\documentclass[prl,reprint,twocolumn,showpacs,preprintnumbers,amsmath,amssymb,nofootinbib]{revtex4-1}
\usepackage{graphicx}

\usepackage{epsfig}
\usepackage{color}
\usepackage{amssymb}
\usepackage{amsmath}
\usepackage{mathrsfs}
\usepackage{hyperref}
\usepackage{multirow}

\begin{document}

\title{Color-Kinematics Duality and Sudakov Form Factor at Five Loops}
\author{Gang Yang}
\affiliation{CAS Key Laboratory of Theoretical Physics, Institute of Theoretical Physics, Chinese Academy of Sciences,  Beijing, 100190, China}
\date{\today}

\begin{abstract}

Using color-kinematics duality, we construct for the first time the full integrand of the five-loop Sudakov form factor in $\mathcal{N}=4$ super-Yang-Mills theory, including non-planar contributions. 
This result also provides a first manifestation of the color-kinematics duality at five loops. 
The integrand is explicitly ultraviolet finite when $D<26/5$, coincident with the known finiteness bound for amplitudes.
If the double-copy method could be applied to the form factor, this would indicate an interesting ultraviolet finiteness bound for $\mathcal{N}=8$ supergravity at five loops.
The result is also expected to provide an essential input for computing the five-loop non-planar cusp anomalous dimension.

\end{abstract}

\pacs{04.65.+e, 11.15.Bt, 11.30.Pb, 11.55.Bq \hspace{1cm}}

\maketitle

\textit{Introduction.}---%
Last decades have seen tremendous progress in the study of scattering amplitudes in quantum field theory and string theory,
in which the maximally supersymmetric $\mathcal{N}=4$ super-Yang-Mills (SYM) theory has been an important testing ground. 
Notably, on-shell unitarity method \cite{Bern:1994zx,Britto:2004nc} and BCFW recursion relation \cite{Britto:2004ap}, initiated in the study of this toy model, have now important applications in computing quantum chromodynamics (QCD) multijet processes at the Large Hadron Collider \cite{Berger:2009zg}. 
In the 't Hooft planar limit, remarkable all loop information has been even obtained for $\mathcal{N}=4$ SYM amplitudes, including the all loop integrand \cite{ArkaniHamed:2010kv} and a non-perturbative interpolation between weak and strong coupling \cite{Basso:2013vsa}.
In comparison, much less is known beyond the planar limit. 
For instance, the cusp anomalous dimension, which is a key quantity for the infrared (IR) singularities of amplitudes \cite{Catani:1998bh}, is known in principle to all orders in planar $\mathcal{N}=4$ SYM \cite{Beisert:2006ez}, but its non-planar correction is unknown even at leading perturbative order.

Promising progress has been made recently through a surprising duality between color and kinematics discovered by Bern, Carrasco and Johansson \cite{Bern:2008qj, Bern:2010ue}, which will be the subject of this Letter. 
This duality indicates a deep connection between the kinematic structure and the color structure in gauge theories. 
Since it involves the full color factors, interlocking the planar and non-planar parts, it offers the promise of transferring the advances of the planar sector to the non-planar sector. 
(See also other intriguing connections between planar and non-planar amplitudes in \cite{Arkani-Hamed:2014via, Bern:2014kca}.)
The duality also allows to construct gravity amplitudes directly from Yang-Mills amplitudes, once the latter are organized to respect the duality.
This is usually referred to as the double copy property \cite{Bern:2010ue, Bern:2010yg}, generalizing the KLT relation \cite{Kawai:1985xq}. Interesting connection between classical solutions in gauge and gravity theories has also been found in \cite{Monteiro:2014cda}.

At tree level, the color-kinematics duality has been proved using monodromy relations in string theory amplitudes \cite{BjerrumBohr:2009rd, Stieberger:2009hq}, or using the BCFW recursion relation directly in field theory \cite{Feng:2010my}. 
However, at loop level the duality is still a conjecture and has only been verified in examples.
Up to four loops, amplitudes respecting the duality have been found in various gauge theories and gravity theories \cite{Bern:2010ue, Carrasco:2011mn, Bern:2012uf, Bjerrum-Bohr:2013iza, Bern:2011rj, Bern:2012uf, Bern:2012cd, Chiodaroli:2013upa, Bern:2014sna, Mafra:2014gja, He:2015wgf}, 
including QCD \cite{Boels:2013bi, Bern:2013yya, Johansson:2015oia, Mastrolia:2015maa, Mogull:2015adi}. 
However, no construction manifesting the duality has been achieved beyond four loops.
The extension to higher loops is a major challenge to understand the duality. 
Via the double copy prescription, such extension would be also essential to resolve the long-standing UV finiteness problem of maximal supergravity at five loops and beyond \cite{Bern:2006kd, Green:2010sp,Bjornsson:2010wm,Beisert:2010jx,Bossard:2011tq, Bern:2012cd, Bern:2014sna}.

In this Letter, we realize the color-kinematics duality for the first time at five loops.
The object we consider is the Sudakov form factor in $\mathcal{N}=4$ SYM, which is an important observable with a gauge-invariant operator in the stress tensor supermultiplet and two on-shell massless states \cite{vanNeerven:1985ja, Gehrmann:2011xn, Boels:2012ew}. 
As the operator is half-BPS, the form factor is protected from ultraviolet (UV) divergences in four dimensions.
In practice, one may simply consider $\langle \phi (p_1) \phi (p_2) | \textrm{Tr} ( \phi^2) |0\rangle$, with all other Sudakov form factors in the multiplet related by supersymmetric Ward identities. 
Since the color-kinematics duality is not generally proved, it is a priori not guaranteed that there exists a five-loop solution that respects the duality. 
In fact, this construction for amplitudes, despite considerable efforts and interest, has not been achieved.
Our result manifests the color-kinematics duality at five loops for the first time, strongly indicating the duality should be true in more general ground.

Sudakov form factor also plays a key role in the study of IR singularities of gauge theories \cite{Mueller:1979ih, Magnea:1990zb}. 
In particular it determines the cusp anomalous dimension. 
The knowledge of its non-planar corrections, which is still unknown, is very important to resolve the full structure of gauge theory IR singularities, see e.g.\ \cite{Gardi:2009qi,Becher:2009qa}.
We expect the new five-loop integrand will provide an essential input towards understanding the non-planar IR singularities, given the tremendous progress on integral techniques (see e.g.\ \cite{Smirnov:2012gma} for review), especially the four-loop form factor integrals \cite{Boels:2015yna, vonManteuffel:2015gxa, Henn:2016men}. 
The form factor we consider is also related to the Higgs production via gluon fusion \cite{Georgi:1977gs}.
We would like to mention that although we consider $\mathcal{N}=4$ SYM, the result is expected to provide the leading transcendentality contribution in QCD \cite{Kotikov:2004er,Brandhuber:2012vm}.

Apart from these, higher loop results also provide key information to understand the explicit UV divergences of gauge and gravity theories. 
The integrand we construct, while satisfying the color-kinematics duality, also manifestly saturates the known finiteness bound for $\mathcal{N}=4$ SYM amplitudes \cite{Bern:1998ug,Howe:2002ui, Bern:2007ct}
\begin{equation}
\label{eq:UV-bound}
D < 4 + {6\over L} \,, \qquad L>1 \,.
\end{equation}
A duality-satisfying construction for amplitudes, via the double copy prescription, would provide crucial input to the UV finiteness problem of $\mathcal{N}=8$ supergravity.
As we will see, our result indicates the UV finiteness of the maximal supergravity at five loops when $D<22/5$, if the double copy of the form factor has a physical meaning in the supergravity. We will discuss more on this in the end.

\textit{Review and result.}---%
The color-kinematics duality conjectures that there exists a cubic (trivalent) graph representation of a general amplitude in gauge theories, such that the kinematic numerators satisfy equations in one-to-one correspondence with Jacobi relations of the color factors \cite{Bern:2008qj, Bern:2010ue}.
The duality applies also to form factors, and explicit constructions have been obtained up to four-loop order \cite{Boels:2012ew}. 

\begin{figure}[t]
\centering
\includegraphics[clip,scale=0.56]{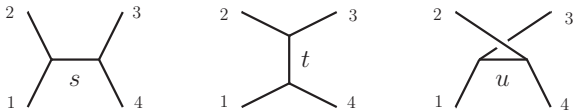}
\caption[a]{Cubic graphs for four-point tree amplitudes.}
\label{fig:treeA4}
\end{figure}

The instructive example is the four-gluon tree amplitude. 
It is always possible to represent the amplitude in terms of three cubic graphs shown in fig.\ \ref{fig:treeA4}, 
\begin{equation}
\label{eq:BCJ-4pt-tree}
{\cal A}^{\rm tree}_4(1, 2, 3, 4) \, = \, {C_s \, N_s \over s} + {C_t\, N_t \over t} + {C_u \, N_u \over u} \,,
\end{equation}
where $C_i$ are color factors as products of structure constants $\tilde f^{abc}$ associated to each cubic vertex. 
The physical information is encoded in the kinematic numerators $N_i$. The color-kinematics duality requires that the numerators should satisfy the Jacobi relation of color factors as
\begin{equation} 
C_s = C_t + C_u  \ \  \Rightarrow \ \  N_s = N_t + N_u \,.
\end{equation}

\begin{figure}[t]
\centering
\includegraphics[clip,scale=0.4]{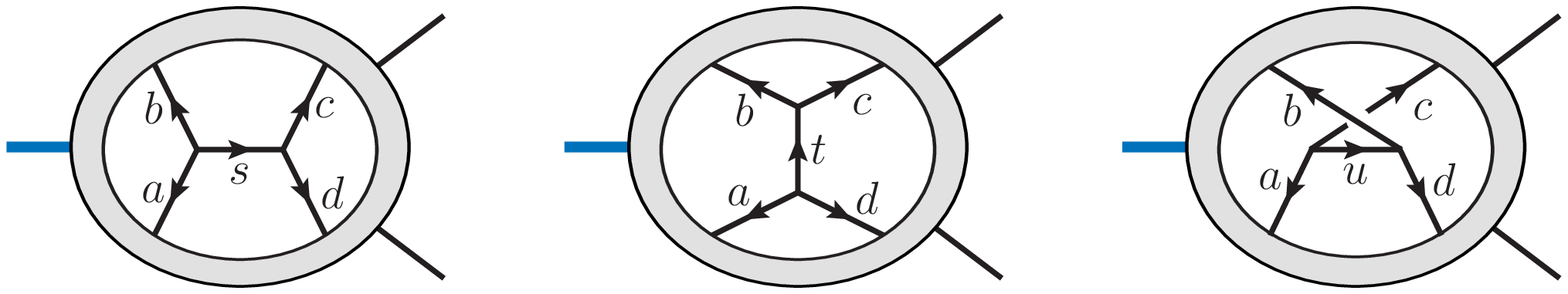}
\caption[a]{Color-kinematics related graphs at loop level.}
\label{fig:loopBCJ}
\end{figure}

While at tree level this has been proved more generally, the duality is conjectured to also hold at loop level.
For any propagator of a trivalent graph (except the internal lines connected to the operator for a form factor), one can take it as in a $s$ channel four-point sub-amplitude, associated with $t$- and $u$-channel graphs, as shown in fig.\ \ref{fig:loopBCJ}.
The duality requires the numerators of the three graphs satisfy the same Jacobi relation for color factors as
\begin{align} 
\label{eq:CK-numerator-relation}
&N_s(\{l_a,l_b,l_s\},\{-l_s,l_c,l_d\}, ...) = \nonumber\\
& \qquad\qquad  N_t(\{l_d,l_a,l_t\},\{-l_t,l_b,l_c\}, ...) \nonumber\\
& \qquad\qquad + N_u(\{l_a,l_c,l_u\},\{-l_u,l_b,l_d\}, ...) \,, 
\end{align}
where $l_i$ label the momenta, $\{l_a,l_b,l_s\}$ specify cubic vertices in the graphs, and the omitted vertices are all identical in the three diagrams. 
Such a relation is understandable if four momenta $l_a, l_b, l_c, l_d$ are on-shell. 
The non-trivial point of the conjectured duality is that it also holds when all propagators are off-shell, as checked so far.
The dual Jacobi equations eq.\ \eqref{eq:CK-numerator-relation} play a central role in our construction.

Before entering the construction, let us give the final five-loop Sudakov form factor in $\mathcal{N}=4$ SYM organized in the following form
\begin{equation}  
\label{eq:F5loop-sum}
{\cal F}_2^{5\textrm{-loop}} = s_{12}^2\, F_2^{\rm tree}  \sum_{\sigma_2} \sum_{i=1}^{306} \int \prod_j^L d^D \ell_j {1\over S_i} \,{C_i \, N_i \over \prod_{\alpha_i} P^2_{\alpha_i}} ,
\end{equation}
where we sum over 306 non-isomorphic cubic graphs. 
The sum over $\sigma_2$ is due to the permutation of external on-shell momenta $p_1$ and $p_2$. 
The symmetry factors $S_i$ remove overcounts from the automorphism symmetries of the graphs. 
Explicit expressions of the numerators $N_i$, color factors $C_i$, symmetry factors $S_i$ and propagator lists $P_{\alpha_i}$ are given in the Supplemental Material \cite{SupplementalMaterial}. 

Below we construct the five-loop result in eq.\ \eqref{eq:F5loop-sum} via color-kinematics duality, together with the constrains from unitarity cuts. 
Reader is referred to  \cite{Bern:2012uf, Boels:2012ew, Carrasco:2015iwa} for the further details of general strategy.

\textit{Five-loop Ansatz.}---%
The starting point is to generate a set of needed cubic graphs. 
For Sudakov form factor, these are graphs with three external legs, two with on-shell momenta $p_1, p_2$ and one with off-shell momentum $q$ associated to the local operator. 
An important simplification for $\mathcal{N}=4$ SYM is that graphs containing tadpole, bubble or triangle  one-loop subgraphs can be excluded, which are known to be valid for the duality-satisfying numerators up to four loops \cite{Bern:2012uf, Boels:2012ew}. 
One-loop sub-triangle is allowed if one of its legs is the external $q$-leg. 
The number of contributed graphs for Sudakov form factor up to five loops is summarized in Table\ \ref{tab:counting}.

\begin{table}[t]
\caption{Number of cubic and master graphs up to five loops.}
\label{tab:counting}
\centering
\begin{tabular}{l | c | c | c | c | c  } 
\hline\hline
$L$ loops  		& \, $L$=1 \,  & \, $L$=2 \, & \, $L$=3 \, & \, $L$=4 \, & \, $L$=5 \,   \cr \hline 
\# of topologies   	&  1 & 2 & 6 & 34 & 306  \cr \hline 
\# of planar masters       	&  1 & 1 & 1 & 2 & 4  \cr \hline \hline
\end{tabular} 
\end{table}

\begin{figure}[t]
\centering
\includegraphics[clip,scale=0.34]{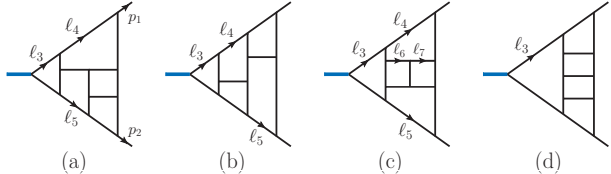}
\caption[a]{Master graphs for the five-loop Sudakov form factor.}
\label{fig:masters}
\end{figure}

The next step is to find a minimal set of {\it master graphs}, from which one can generate all other graphs using the dual Jacobi relations eq.\ \eqref{eq:CK-numerator-relation}.
Since the system of Jacobi equations are highly constraining, the number of master graphs is usually very small. 
The choice of master graphs is not unique. 
For convenience we choose all master graphs to be planar, as shown in fig.\ \ref{fig:masters}. 
Note that it is possible to replace two of the planar master graphs by a non-planar one, which reduces the number of masters to be three.
A counting of planar master graphs up to five loops is given in Table \ref{tab:counting}.

The third step is to make an ansatz for the numerators of master graphs. 
After an overall factor $s_{12}^2 F_2^{\rm tree}$ is factorized as in eq.\ \eqref{eq:F5loop-sum}, the remaining numerators should be polynomials of degree six in the loop and external momentum. 
As in previous observation at four loops, we impose the power counting conditions as follows: 
a one-loop $n$-gon subgraph carries no more than $n-4$ powers of loop momentum for that loop, which is a consequence of supersymmetry; 
and if the $n$-gon subgraph is attached to the $q$-leg, it carries no more than $n-3$ powers of the loop momentum, which is related to the fact that the operator is half-BPS. 
For example for master graph fig.\ \ref{fig:masters}(c), the numerator should be no more than linear in the loop momenta $\ell_4, \ell_5$ and at most quadratic in $\ell_3$. 
This gives an ansatz of the sum over 77 Lorentz products of $\ell_i, p_i$ with 77 parameters.
With numerators of other three master graphs, we have in total $162$ parameters. 

Choosing a proper set of dual Jacobi relations eq.\ \eqref{eq:CK-numerator-relation}, we can express all other numerators in terms of the four master ansatz-numerators.
Note that we have used the no-triangle and power counting properties of $\mathcal{N}=4$ SYM to simplify the ansatz. Whether such an ansatz should be sufficient at five loops is 
a priori not clear. If not, one would need to relax some conditions and enlarge the ansatz space. 
As we will see, the above ansatz turns out to be sufficient for the construction.

\textit{Solving ansatz and checks.}---%
To fix the parameters, we first demand that each numerator respects the automorphism symmetries of the graph.
Take master graph fig.\ \ref{fig:masters}(c) as an example. 
There is one automorphism symmetry, which constrains the numerator to be invariant under the transformation
\begin{align}
\{ & \ell_3 \rightarrow p_1+p_2 - \ell_3, \quad \ell_4 \leftrightarrow \ell_5,  \nonumber\\
&
 \ell_6 \rightarrow p_1 + p_2 -\ell_4-\ell_5 - \ell_6,  \nonumber\\
& \ell_7 \rightarrow  p_1 + p_2 -\ell_4-\ell_5 - \ell_7 \} .
\end{align}
Applying symmetry constraints for all cubic graphs fixes 115 parameters.

Next we employ physical constraints via unitarity cuts
\begin{equation}
\label{eq:unitarity-cut} 
{\rm cut}(\sum\textrm{cubic graphs})  = \sum_{\rm states} \, F^{\rm tree} \prod_I A_I^{\rm tree}  ,
\end{equation}
where the cut integrand of the cubic graph ansatz should be equal to the product of physical tree quantites.
We start with some simple maximal cuts. It is easy to write the rung-rule numerators for all master graphs \cite{Bern:1997nh}. 
For instance for master graph fig.\ \ref{fig:masters}(b),
\begin{align}
N^{\rm (b)}_\textrm{rung-rule} & = s_{12} (\ell_3 +\ell_5 - p_1 - p_2)^2 (\ell_4 - p_1)^2 \\
&  + (\ell_3 - p_1)^2 (\ell_4 - p_1 - p_2)^2 (\ell_5 - p_1 - p_2)^2 \nonumber\,.
\end{align}
Under maximal cuts, the ansatz-numerator and the rung-rule-numerator must be equal.
Applying this for four master graphs fixes further 27 parameters. 

Since so far we only use a subset of dual Jacobi relations to generate all numerators, we should check if all other Jacobi relations are satisfied. 
This provides 10 more constraints. 
We are thus left with only 10 parameters.

\begin{figure}[t]
\centering
\includegraphics[clip,scale=0.533]{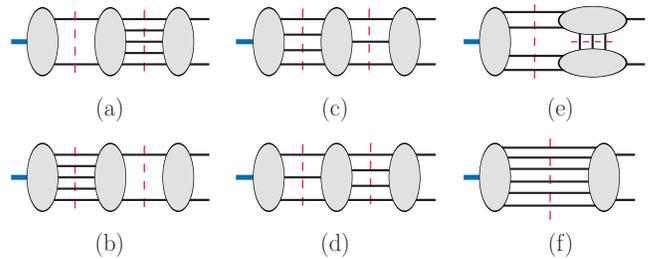}
\caption[a]{Non-trivial unitarity cuts of five-loop form factor. Each blob denotes a tree amplitude or form factor.}
\label{fig:F2_5loop_multicuts}
\end{figure}

Given the small number of parameters, we are ready to consider more general cuts as given in fig.\ \ref{fig:F2_5loop_multicuts}. 
We find that applying cuts (a)-(d) in fig.\ \ref{fig:F2_5loop_multicuts} fixes 7 parameters,  leaving an integrand with only 3 unfixed parameters. 
Further checks of cuts (e) and (f) in fig.\ \ref{fig:F2_5loop_multicuts} show that the 3-parameter integrand automatically satisfies these cuts.

The check of cut (f) is one of the most involving cuts, so we provide some details.
From the r.h.s.\ of eq.\ \eqref{eq:unitarity-cut}, it is the product of a six-point form factor and an eight-point amplitude. 
One needs to sum over all possible states of the cut legs. 
This includes summing over all helicity configurations including non-trivial non-MHV tree results up to N$^4$MHV.
Such tree amplitudes and form factors, including the summing of states, can be computed with MHV rule method \cite{Cachazo:2004kj, Elvang:2008na, Bern:2009xq}. 
On the other hand, from the ansatz side i.e.\ l.h.s.\ of eq.\ \eqref{eq:unitarity-cut}, it involves 79 trivalent topologies which generate more than a thousand cut diagrams. 
These two highly non-trivial expressions, computed from different origins, match perfectly with each other.

\begin{figure}[t]
\includegraphics[clip,scale=0.72]{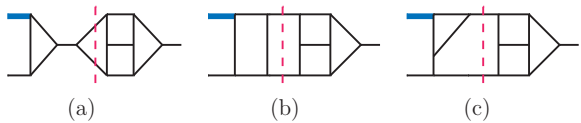}
\caption[a]{Unitarity check of subtle 1PR graphs.}
\label{fig:subtlegraphs}
\end{figure}

One subtle unitarity check is related to the graphs as fig.\ \ref{fig:subtlegraphs}(a). Such a graph is apparently not well-defined: it is one-particle-reducible (1PR) and contains a two-point subgraph on an on-shell leg, giving a divergent propagator $1/p_i^2 = 1/0$. However, the numerator could also contain a factor $p_i^2$ which cancels the divergent propagator. Similar subtlety has indeed appeared in the case of four-loop four-point amplitude \cite{Bern:2012uf}. So one needs to check if they contribute or not. This can be fixed by studying the cuts shown in fig.\ \ref{fig:subtlegraphs}. We found that the (b) and (c) type graphs always cancel among themselves, which means the (a) type graphs have zero numerators.

\begin{figure}[t]
\centering
\includegraphics[clip,scale=0.63]{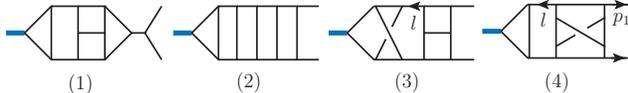}
\caption[a]{A fish-type graph and three other sample graphs of five-loop Sudakov form factor.}
\label{fig:samplegraphs}
\end{figure}

An interesting feature comes from the `fish graphs' that contain a two-point tree leg, as shown in fig.\ \ref{fig:samplegraphs}(1). 
Their color factors are zero, so they do not contribute to the final form factor. 
However, they do have non-zero numerators to satisfy the color-kinematics duality and unitarity constraints.
The tree propagator seems to indicate the graph is 1PR. 
However, the numerator can be proportional to $s_{12}$ which cancels the propagator. 
Imposing this cancellation condition fixes two of the remaining three parameters, although they are not required by unitarity.

While extensive checks have been done for our result, a complete proof of the result would require non-planar as well as $D$-dimensional unitarity checks. Given the fact that color-kinematics duality is independent of the spacetime dimension and also interlocks the non-planar parts from planar parts, we believe our result is complete, which is indeed the case for previous  constructions up to four loops.
We leave the complete check for future work.

Let us give some simple numerator examples. For the last three graphs shown in fig.\ \ref{fig:samplegraphs}, we have
\begin{align}
N^{(2)} = & s_{12}^3 \,, \quad N^{(3)} = s_{12}^2 l\cdot (p_1-p_2) - {1\over2} s_{12}^3  \,, \\
N^{(4)} = & {2\over9} s_{12} \Big[ 11 (l\cdot p_1)^2 + 11(l\cdot p_2)^2 + 40 (l\cdot p_1)(l\cdot p_2)   \nonumber\\
& + \big(-2l^2 + 18 l\cdot p_1 + 9 l \cdot p_2 + {7\over4} s_{12} \big)s_{12} \Big] \,.
\end{align}

\textit{Summary and discussion.}---%
The final five-loop form factor result is given in eq.\ \eqref{eq:F5loop-sum} in terms of 306 non-isomorphic graphs. 
They include 60 graphs that have zero color factors, and 26 of them are fish graphs. 
We keep them since they do have non-trivial numerators.
The five-loop integrand contains an unfixed parameter (or three if we do not require the $1/s_{12}$ pole to be cancelled by the numerator in fish graphs). 
This also happens in the four-loop case and it was shown that the free parameter drops out after the integration-by-part reduction \cite{Boels:2015yna}. 
It is reasonable to expect similar things happen at five loops. 
It would be interesting to study the duality for generic form factors in $\mathcal{N}=4$ SYM (see \cite{Wilhelm:2016izi} for recent reviews) and in QCD.

Let us comment on the UV behavior of the result. The five-loop cubic graphs have 15 propagators, and the numerators are at most quartic in the loop momenta. 
Therefore, the integrand is manifestly UV finite when $D<26/5$ which saturates the finiteness bound given in eq.\ \eqref{eq:UV-bound} for $\mathcal{N}=4$ SYM  amplitudes \cite{Bern:1998ug,Howe:2002ui, Bern:2007ct}. Note the bound is also saturated for Sudakov form factor at three and four loops \cite{Gehrmann:2011xn, Boels:2012ew}.
We mention that the numerators of fish graphs like fig.\ \ref{fig:samplegraphs}(1) contain up to third power of loop momentum, which could break the bound if their color factors are not zero. 
If the double copy of the form factor has a physical meaning, our result indicates the UV finiteness of five-loop maximal supergravity when $D<22/5$.
Notably, studies based on symmetry arguments in maximal supergravity have lead to the expectations of a seven-loop divergence in $D=4$ and a five-loop divergence in $D=24/5$ for amplitudes \cite{Green:2010sp,Bjornsson:2010wm,Beisert:2010jx,Bossard:2011tq}.
While our result seems to indicate a lower bound, there could well be some enhanced cancellations that are not visible from the power counting of the local integrand form, as shown in \cite{Bern:2006kd, Bern:2012cd, Bern:2014sna}. 
It would be very interesting to check this by explicit integral computations.
While the physical interpretation of the double copy of form factors is still not clear, mainly due to the lack of local observables in gravity, it would be interesting to study this further and see if we can use form factors to shed new light on the UV property of gravity.

Finally, we point out that obtaining integrands with small power of loop momentum in the numerator are also essential for simplifying reduction and integration. 
We expect our result with full non-planar contribution will provide a promising starting point for the computation of the unknown non-planar cusp anomalous dimension at five loops in the future.

We thank Rutger Boels for collaboration on related topics and many discussions. 
We are grateful to Zvi Bern, John Joseph Carrasco, Henrik Johansson, and Radu Roiban for valuable correspondence and comments. 
It is a pleasure to thank Bo Feng, Song He, Johannes Henn, Yu-tin Huang, Jian-Ping Ma and Congkao Wen for discussions. 
We also thank ``Strings 2016" where part of this work was presented. 
This work is supported in part by the Chinese Academy of Sciences (CAS) Hundred-Talent Program and by the Key Research Program of Frontier Sciences of CAS (Grant No. QYZDB-SSW-SYS014).

\bibliographystyle{apsrev4-1}

\end{document}